\documentclass[english,prl,twocolumn,aps]{revtex4}
\usepackage[T1]{fontenc}
\usepackage[latin1]{inputenc}
\usepackage{graphicx}

\makeatletter

\makeatletter

\usepackage{soul}

\makeatother

\usepackage{babel}
\makeatother

\begin{document}

\title{The effect of nonlinearity on adiabatic evolution of light}

\author{Y. Lahini$^{1}$, F. Pozzi$^{2}$, M. Sorel$^{2}$, R. Morandotti$^{3}$,
D. N. Christodoulides$^{4}$ and Y. Silberberg$^{1}$}

\affiliation{$^{1}$Department of Physics of Complex Systems, the Weizmann Institute
of Science, Rehovot, Israel}

\email{yoav.lahini@weizmann.ac.il}

\affiliation{$^{2}$Department of Electrical and Electronic Engineering, University
of Glasgow, Glasgow, Scotland}

\affiliation{$^{3}$Institut National de la Recherche Scientifique, Universit$\acute{e}$
du Qu$\acute{e}$bec, Varennes, Qu$\acute{e}$bec, Canada}

\affiliation{$^{4}$CREOL/College of Optics, University of Central Florida, Orlando,
Florida, USA}

\begin{abstract}
We investigate the effect of nonlinearity in a system described by
an adiabatically evolving Hamiltonian. Experiments are conducted in
a three-core waveguide structure that is adiabatically varying with
distance, in analogy to the STIRAP process in atomic physics. In the
linear regime, the system exhibits an adiabatic power transfer between
two waveguides which are not directly coupled, with negligible power
recorded in the intermediate coupling waveguide. In the presence of
nonlinearity the behavior of this configuration is drastically altered
and the adiabatic light passage is found to critically depend on the
excitation power. We show how this effect is related to the destruction
of the dark state formed in the STIRAP configuration.
\end{abstract}


\maketitle The adiabatic theorem describes one of the most
powerful concepts in quantum physics\cite{QM}. It states that if
the parameters of a quantum system evolve slowly enough in time,
the associated initial eigenstates will be preserved, and there
will be no exchange of energy between them. This well studied
theorem finds wide applications in diverse areas of science
ranging from molecular physics to quantum field theory, from
chemistry to nuclear physics. A close reexamination of the
adiabatic principles led to the discovery of Berry's geometric
phase\cite{Berry} - known to occur ubiquitously in many processes
in nature\cite{Berry2}. Quite recently, quantum adiabatic methods
were suggested as a basis for a new class of algorithms meant to
address NP-complete problems within the framework of quantum
computing\cite{Farhi}. In addition, techniques exploiting an
adiabatic passage provide practical approaches in achieving nearly
complete population transfer between two quantum
states\cite{STIRAP1,STIRAP2,CM,Vardi}. One such example of
coherent adiabatic excitation is stimulated Raman adiabatic
passage (STIRAP) that makes use of two appropriately prepared
laser pulses in order to couple two non-degenerate metastable
states via an intermediate level. Remarkably this can be achieved
without any appreciable excitation of the intermediate
state\cite{STIRAP1,STIRAP2,Vitanov}.

One of the underlying - and sometimes limiting- assumptions of the
adiabatic theorem is the presumed intrinsic linearity of the
system, a condition that is often not met under actual
experimental conditions. For example, nonlinearity comes into play
in various adiabatically evolving systems such as Bose-Einstein
condensates in slowly varying potentials or
fields\cite{BEC,BEC2,Pu,Pu2} and nonlinear optical
processes\cite{Silberberg,Assanto}. Of course, the question
naturally arises on how nonlinear effects may influence such
adiabatic transfer processes\cite{Niu,Pu,Pu2,BEC2,Polkovnikov} -
an aspect that has so far eluded experimental observation.

In this letter we consider the influence of nonlinearity in
systems described by an adiabatically evolving Hamiltonian.
Experiments are conducted in a system of coupled optical
waveguides, in which the distance between channels changes slowly
along the propagation axis. Nonlinear optical waveguides,
described by the nonlinear Schr$\ddot{o}$dinger equation, allow
one to take a simple and direct experimental look at the interplay
between adiabatic evolution and nonlinearity. In addition they
provide a direct analogy with various other quantum processes.
These include time-dependent quantum effects in atomic physics,
Bose-Einstein condensates in time varying traps and time dependent
quantum-well potentials - all described in different regimes by
the same evolution equations presented here. As an example, we use
a three-waveguide structure that reproduces the STIRAP process in
atomic physics\cite{Longhi2}. In the linear regime, the system
exhibits a complete and irreversible power transfer between two
waveguides that are not directly coupled, via an intermediate
channel. Remarkably, this intermediate waveguide carries no
significant field amplitude during the power exchange. In the
nonlinear regime, the adiabatic light passage is found to
critically depend on the excitation power levels. We show how this
effect is related to the destruction of a dark state formed in the
STIRAP configuration\cite{Pu2}.

\begin{figure}
\includegraphics[clip,width=1\columnwidth]{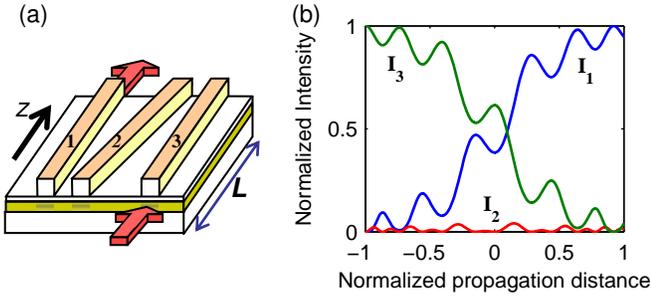}

\caption{(color online). (a) A schematic view of the STIRAP sample. The relative
distance between the coupled waveguides (denoted 1,2 and 3) changes
slowly along the propagation axis z, resulting in slowly changing
couplings rates between the waveguides. (b) Adiabatic light passage
as calculated from Eq. (2) for a 3-core structure with $\alpha=66m^{-1},L=3cm,\kappa=600m^{-1}$(see
text for definitions). The intensity in every channel is plotted as
a function of normalized distance.}

\end{figure}

Consider a system of three single-mode, evanescently coupled nonlinear
waveguides (denoted as 1, 2 and 3, see Fig. 1a). The waveguides are
identical in shape and have a constant width along the propagation
direction, $z$. However, the distances between the waveguides vary
along the propagation. Waveguides 1 and 3 are parallel to each other,
while waveguide 2 is oblique; it is closer to waveguide 1 at $z=0$,
and closer to waveguide 3 at $z=L$, where $L$ is the sample length
(see Fig. 1a). As a consequence, the coupling rates between the waveguides
vary slowly along the propagation. At $z=0$ the coupling between
waveguides 1 and 2 $(C_{12})$ is strong, while the coupling between
waveguide 2 and 3 $(C_{23})$ is weak. At the output of the system
(at $z=L$) the situation is reversed, i.e. $C_{23}>C_{12}$. The
coupling between waveguides 1 and 3 is practically zero in this configuration.

The evolution of the modal amplitudes in these three waveguides can
be described by the following set of coupled discrete nonlinear Schr$\ddot{o}$dinger
equations: \begin{equation}
i\frac{\partial{E_{n}}}{\partial{z}}+\beta_{n}E_{n}+\sum_{m}{C_{n,m}(z)E_{m}}+\Gamma|E_{n}|^{2}E_{n}=0\label{eq:1}\end{equation}

where $n=1,2,3$, $E_{n}$ is the wave amplitude in waveguide n, $\beta_{n}$
is the longitudinal wavevector (propagation constant) for the mode
or bound state in waveguide $n$ and the summation is carried out
on nearest-neighbors. The last term in Eq.(1) accounts for the nonlinear
dependence of the on-site wavevector $\beta$, where $\Gamma$ is
associated with the Kerr nonlinear coefficient of the waveguide structure.
This term is important only in the nonlinear regime and can be neglected
at low light power levels.

In the linear limit, the description of this system by Eq.(1)
carries a perfect analogy to the STIRAP process, first described
in the framework of atomic physics\cite{STIRAP1,STIRAP2}. This
surprising process leads to a complete transfer of population
between two atomic levels for which a direct transition is
forbidden, via a third level. However, in the adiabatic limit the
intermediate level is never populated during the
process\cite{STIRAP2}. Indeed, the equations used to describe the
STIRAP effect in atomic physics are identical, under the rotating
wave approximation, to Eq.(1) in the linear limit. In this analogy
z replaces time, the amplitude in each waveguide corresponds to
the amplitude in each atomic level and the coupling between the
waveguides plays the role of the Rabbi coupling of the atomic
energy levels caused by resonant electromagnetic radiation.
Identical values of the parameter$\beta$ for coupled waveguides
represent zero detuning of the electromagnetic radiation from the
level spacing. A linear STIRAP scheme was recently suggested in an
optical system using a different analogy that required an imprint
of periodic gratings or bending of the waveguides along the
propagation axis, to introduce coupling between dissimilar
waveguides\cite{Longhi1,Orenstein}. An implementation using
identical waveguides and a simple geometry similar to the one
discussed here was proposed by Paspalakis\cite{Paspalakis},
and recently implemented in the linear regime by Longhi and coworkers\cite{Longhi2}.%
\begin{figure}
\includegraphics[clip,width=1\columnwidth]{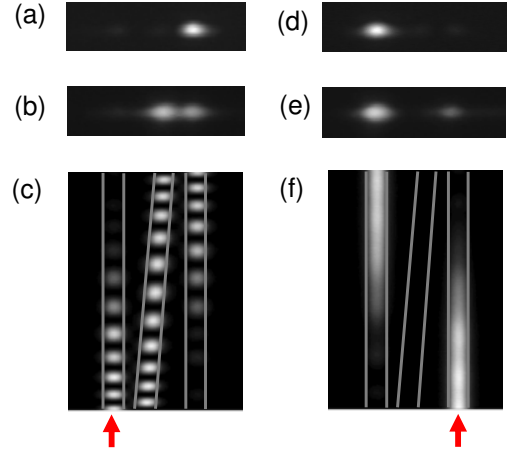}

\caption{(color online). Adiabatic passage in the STIRAP sample. (a) Measurement
of the output light distribution when light is injected into waveguide
1. After the adiabatic sweep, the light emerges from waveguide 3.
(b) The same experiment in a truncated sample, showing that during
the adiabatic sweep, there is significant intensity in waveguide 2.
(c) BPM simulations of the propagation. (d)-(f) The same as (a)-(c),
when light is injected to waveguide 3 and emerges from waveguide 1.
In this case, during the adiabatic sweep the intensity in waveguide
2 is negligibly small.}

\end{figure}

To theoretically analyze the linear response ($\gamma=0$) of the
system shown in Fig. 1a we recall that the coupling coefficient
between two evanescently coupled waveguides varies exponentially
with the separation distance\cite{Yariv}. As a result, for a
structure of length L, the two coupling constants are found to
vary according to $C_{12}(z)=\kappa\cdot exp[-\alpha(z-L/2)]$ and
$C_{23}(z)=\kappa\cdot exp[\alpha(z-L/2)]$, where $\kappa$ is the
coupling strength in the middle of the structure $(z=L/2)$ and
$\alpha$ is a slow adiabatic parameter related to the slope of
waveguide $2$, that is $\gamma\equiv\alpha/\kappa\ll1$. If at the
input of this system, the third waveguide is excited, i.e.
$E_{3}(0)=1$ , then by employing WKB expansion methods one can
show that to a very good approximation the field in the first
waveguide evolves according to: \begin{equation}
E_{1}(z)e^{-i\beta
z}=\frac{A\sqrt{1+e^{4t_{0}}}}{\sqrt{1+e^{-4t}}}-\frac{A\sqrt{1+e^{-4t_{0}}}}{\cos\phi\sqrt{1+e^{4t}}}\cos[Q(t)+\phi]\end{equation}

In Eq.(2), $A^{-1}=[-4\gamma^{2}-2\gamma^{2}\tanh(2t_{0})-2\cosh(2t_{0})]$,
$t_{0}=\alpha L/2,$ $t=\alpha(z-(L/2)),$ $-t_{0}\leq t\leq t_{0}$,
$\tanh(\phi)=-\gamma(2/\cosh(2t_{0}))^{1/2}$, and $Q(t)$ is a phase
function. $E_{2}$ and $E_{3}$ are obtained by plugging Eq. (2) into
Eq. (1), and using the conservation law $|E_{3}|^{2}=1-|E_{1}|^{2}-|E_{2}|^{2}$.
Fig. 1b shows the evolution of the intensities $I_{n}=|E_{n}|^{2}$
in a 3-core adiabatic system with parameter values very close to those
used in our experiments, as obtained from the analytical expressions
of Eq. (2). The numerical results are not shown here since they are
very close to those already depicted. As clearly shown in Fig. 1b,
the power adiabatically leaves channel 3 and eventually populates
channel 1, with very little energy remaining in the intermediate waveguide.
This is in perfect analogy to the STIRAP process. The first term on
the right of Eq.(2) is primarily responsible for this adiabatic transition
whereas the second one describes the oscillatory component in Fig.
1b.

The waveguide triplet used in our experiment was fabricated on an
AlGaAs substrate, using standard photolithography
techniques\cite{GSS}. The waveguides have a width of 3 $\mu m$,
and the sample length is $L$=18mm. The edge-to-edge distance
between waveguide 1 and 2 is 2 $\mu m$ at z=0, and 7 $\mu m$ at
z=L, while the distance between waveguide 1 and 3 is fixed at 12
$\mu m$. This yields a coupling of ~2500 $m^{-1}$ between
waveguide 2 and 3 at z=L=18 mm and a coupling of ~250 $m^{-1}$
between waveguide 2 and 3 at z=0, while the coupling between the
waveguides is ~790 $m^{-1}$ at z=L/2. A second sample with similar
parameters was fabricated, and was truncated to enable observation
of the amplitude in the waveguides before the full sweep is
achieved. In the experiments presented below, light is injected
into one of the waveguides in the structure at z=0, propagates
along the sample and is measured at the sample output.
Nonlinearity is introduced by increasing the power of the input
beam. A full description of the
experimental setup can be found elsewhere\cite{GSS}. %
\begin{figure}
\includegraphics[clip,width=0.9\columnwidth]{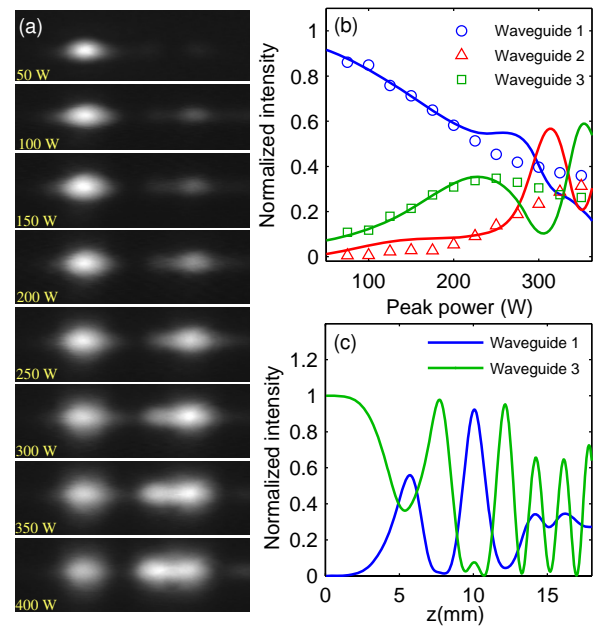}

\caption{(color online). The effect of nonlinearity on adiabatic passage. (a)
Measurements of the output light distribution as in Fig. 2d, at different
input intensities. (b) comparison between the experimental results
(markers) and numerical calculations (lines, see text). (c) Numerical
calculations of the intensity distribution in the sample along the
propagation, for an input power of 350W.}

\end{figure}

Fig. 2 shows the result of experiments done at low powers. When the
input beam is launched into waveguide 1 (Fig. 2a), the output light
emerges from waveguide 3. However, a similar experiment done in the
truncated sample (Fig. 2b), reveals that waveguide 2 carries a significant
field amplitude during the power exchange between waveguide 1 and
waveguide 3. This is also illustrated in the BPM simulation shown
in Fig. 2c. On the other hand, when light is initially injected into
waveguide 3, it emerges from waveguide 1 as shown in Fig. 2d, yet
the truncated sample shows that in this case the intesity in waveguide
2 is negligible during the process (Fig. 2e). This result is corroborated
by the BPM simulation, as shown in Fig. 2f. Even though the coupling
between waveguide 3 and waveguide 1 is zero, power is fully exchanged
between them during the adiabatic sweep, with only minimal excitation
of waveguide 2.

We now turn to the effect of nonlinear perturbations on the
adiabatic passage described above. For this purpose we again
launched light into waveguide 3, and measured the output light
distribution as a function of the input beam power. The results of
this experiment are presented in Fig. 3a. These results show that
the presence of nonlinearity reduces the efficiency of the
adiabatic passage, even at relatively weak powers. The
experimentally measured light distribution at the output are
compared to BPM numerical results in Fig. 3b, taking into account
corrections due to dispersion and cross-phase modulation
effects\cite{XPM}. The experimental and numerical results show
good agreement in the weak nonlinear regime, while at higher
powers the experiment deviates from the theory, probably due to
nonlinear absorption effects. Fig. 3c shows an example of the
calculated evolution of the intensities in waveguides 1 and 3
along the propagation in the nonlinear regime (power of 350W).
This figure should be compared with the linear dynamics in Fig 1b.

These results are compatible with previous theoretical predictions
that considered the mean-field dynamics of a Bose-Einstein
condensate in a time dependent triple-well trap\cite{BEC2}. The
authors have shown that the adiabatic passage should break down
when the magnitude of the nonlinear parameter $\Gamma$ exceeds
that of the detuning between levels. In the optical analogue,
detuning is introduced when the waveguides have different
propagation parameters $\beta.$ In the configuration used here all
three waveguides are identical, which corresponds to zero
detuning, hence the adiabatic passage is expected to break down
even for very weak nonlinearity.

\begin{figure}
\includegraphics[clip,width=0.9\columnwidth]{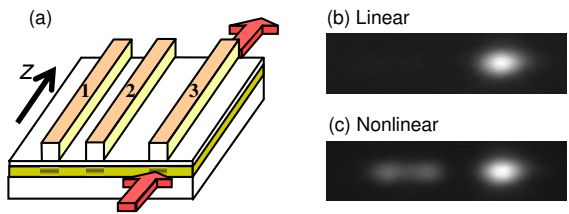}

\caption{(color online). (a) A schematic view of the sample used to probe the
dark state. (b) Formation of the dark state in the linear regime.
Light is injected to waveguide 3 and remains trapped there, despite
the coupling between waveguide 3 and waveguide 2 (see text). (c) Partial
destruction of the dark state by nonlinearity. }

\end{figure}

The STIRAP effect relyes on the existence of a dark eignstate of
the system, a phenomenon known as Coherent Population Trapping
(CPT)\cite{STIRAP2,Vitanov}. It has been theoretically shown that
dynamical level shifts induced by nonlinearity can affect the
resonance condition that leads to the CPT state, hence reducing
the efficiency of STIRAP\cite{Pu2}. To demonstrate this effect in
our system, we consider the configuration presented in Fig. 4a
which is identical to the configuration of the STIRAP sample at
z=0, but with no variation of the couplings along the z direction.
Waveguides 3 and 2 are weakly coupled, therefore light injected
into waveguide 3 is expected to tunnel along the propagation to
waveguide 2. However, the strong coupling between waveguide 2 and
waveguide 1 results in two new modes with propagation constants
that are spaced symmetrically around that of the third guide. This
leads to a sharp resonance that eliminates the tunneling, and
therefore light that is injected into waveguide 3 remains trapped
in that waveguide. The formation of this dark state is
experimentally demonstrated in Fig. 4b. When nonlinearity is
introduced by increasing the input power (300W), the eigenvalue of
the mode in waveguide 3 is shifted and the resonance condition is
no longer satisfied. As a result the dark state is destroyed and
tunneling out of waveguide 3 is partially recovered (Fig. 4c).
Since the STIRAP effect is based on the evolution of this dark
state, this also explains the sensitivity of STIRAP to
nonlinearity. Is it interesting to note that even though the level
detuning due to nonlinearity can in principle be compensated by
the sample design, the dark state may still be dynamically
unstable\cite{Pu2}.

In summary, using coupled nonlinear optical waveguides we have investigated
the effect of nonlinearity on an adiabatic process - an optical analogue
of the STIRAP process. In the nonlinear regime, we found that even
weak nonlinearity is enough to impair the efficiency of STIRAP. This
was explained by the destruction of the dark state formed in the STIRAP
configuration.

The approach presented here can be extended to more complex structures,
implementing a variety of slowly-varying photonic potentials and giving
rise to new nonlinear effects. Waveguide lattices can be used to adiabatically
introduce changes in the dispersion relation, for example by opening
or shifting gaps in the spectrum or by introducing disorder, offering
a new experimental playground for the study of the interplay between
nonlinearity and adiabaticity.

This work was supported by the German-Israeli Project Cooperation
(DIP), NSERC and CIPI (Canada), and EPRSC (UK). YL is supported by
the Adams Fellowship of the Israel Academy of Sciences and Humanities.
We thank O. Katz for help with the numerical simulations and H. Suchowski
for useful discussions.

\end{document}